\documentclass[12pt]{iopart}
\usepackage{amsfonts}
\usepackage{mathrsfs}
\usepackage{amssymb}
\begin{document}
\renewcommand{\theequation}{\arabic{section}.\arabic{equation}}
\title[]{Bethe States of the integrable spin-$s$ chain with generic open boundaries}
\author{Lijun Yang$^1$, Xin Zhang$^1$, Junpeng Cao$^{1,2}$,
Wen-Li Yang$^{3,4}$\footnote{Corresponding author:
wlyang@nwu.edu.cn}, Kangjie Shi$^3$, Yupeng
Wang$^{1,2}$\footnote{Corresponding author: yupeng@iphy.ac.cn}}
\address{$^1$Beijing National
Laboratory for Condensed Matter Physics, Institute of Physics,
Chinese Academy of Sciences, Beijing 100190, China}
\address{$^2$Collaborative Innovation Center of Quantum Matter, Beijing, China}
\address{$^3$Institute of Modern Physics, Northwest University, Xian 710069, China}
\address{$^4$Beijing Center for Mathematics and Information Interdisciplinary Sciences, Beijing, 100048, China}
\begin{abstract}
Based on the inhomogeneous $T-Q$ relation and the associated Bethe
Ansatz equations obtained via the off-diagonal Bethe Ansatz, we
construct the Bethe-type eigenstates of the $SU(2)$-invariant
spin-$s$ chain with generic non-diagonal boundaries by employing
certain orthogonal basis of the Hilbert space.
\end{abstract}

\vspace{2pc}

\maketitle

\section{Introduction}
The $SU(2)$-invariant spin-$s$ Heisenberg chain has attracted great
attention since its close relationship to the
Wess-Zumino-Novikov-Witten (WZNW) models
\cite{Wess,Novikov,Witten,Thomale} and low-dimensional
super-symmetric quantum field theory \cite{Ahn,Inami,Nepomechie,
Bajnok}. With the fusion techniques
\cite{Kulish1,Kulish2,Kulish3,Kirillov1,Kirillov2}, the integrable
high spin model can be constructed from the fundamental $s=1/2$
representation of the Yang-Baxter equation \cite{Yang,Baxter}. The
model with periodic
\cite{Zamolodchikov,Babujian1,Takhtajan,Babujian2}, anti-periodic
\cite{nicc15} and diagonal open boundaries \cite{Mezincescu,Fireman,
Doikou} has been extensively studied. However, the story for the
spin chains with generic non-diagonal boundaries is quite different
even their integrabilities were known \cite{sklya} for a long time.
For the spin-$\frac{1}{2}$ case, the exact solution was first given
in \cite{Cao2} by the off-diagonal Bethe Ansatz method (ODBA)
\cite{Cao2,Cao1,Cao3,Cao4} (for comprehensive introduction, see
\cite{Book}). It is remarked that some other methods such as the
q-Onsager algebra method \cite{o1,o2,Bas07,o3,o4,o5,o6}, the
separation of variables (SoV) method
\cite{nicc13-1,nicc13-2,nicc13-3,Niccoli,nicc14-1,nicc14-2} and the
modified algebraic Bethe ansatz method
\cite{Belliard,Belliard1,Belliard2,Belliard2-1} were also used to
approach the spin-$\frac{1}{2}$ chain with generic integrable
boundary conditions. We should note that the spin-$\frac{1}{2}$
chain with triangular boundary reflection matrix was studied by
Belliard, Cramp\'{e} and Ragoucy \cite{t1} and later by Pimenta and
Lima-Santos \cite{t2}. Ribeiro, Martins and Galleas obtained the
exact solution of the $SU(N)$-invariant high spin chain with generic
toroidal boundary conditions \cite{t3}. For the
$SU(2)$-invariant spin-$s$ chains (with generic $s$), the exact
solutions for the non-diagonal boundaries were previously known only
for some special cases \cite{clsw,Frappat, Murgan1,Murgan2,
Baiyasi,martins}. Until very recently, exact spectrum of the model
with generic boundary conditions was derived \cite{Cao5} in terms of
an inhomogeneous $T-Q$ relation via the ODBA. However, its
eigenstates are still missing.

Up to now Bethe states, which have well-defined homogeneous limits,
of integrable models with generic open boundaries are only known for
few cases \cite{Belliard,Belliard2,Belliard2-1,Cao6,Cao7}. A
remarkable fact is that the method proposed in \cite{Cao6,Cao7}
allows us to retrieve the eigenstates based on the inhomogeneous
$T-Q$ relations obtained from the ODBA in a systematic way. In this
paper, we adopt this method to derive the Bethe-type eigenstates of
the integrable spin-$s$ chain with generic non-diagonal boundaries.

The paper is organized as follows. In sections 2, we
briefly review the fusion procedure and the ODBA solutions of
the integrable spin-$s$ chain with generic open boundary condition.
In section 3, we introduce a gauge transformation and commutation
relations, which are quite useful in the following derivations. Section 4 is devoted to the
construction of an orthogonal basis of the Hilbert space. In section 5, we show that the scalar product between an eigenstate and a basis vector can be expressed in terms of the
corresponding eigenvalues. A useful inner
product is calculated in section 6. Section 7 is devoted to the
construction of the Bethe-type eigenstates. We summarize our
results in section 8.

\section{The model and its spectrum}
\setcounter{equation}{0}

The $R$-matrix of the spin-$s$ Heisenberg spin chain is \cite{Kulish1,Kulish2,Kulish3}
\begin{equation}
R_{1,2}^{(s,s)}(u)=\prod_{j=1}^{2s}(u-j\eta)\sum_{l=0}^{2s}\prod_{k=1}^l\frac{u+k\eta}{u-k\eta}P_{1,2}^{(l)},
\label{r111}
\end{equation}
where $u$ is the spectral parameter, $\eta$ is the crossing
parameter and $P_{1,2}^{(l)}$ projects the tensor space of two
spin-s into the irreducible subspace of spin-$l$
\begin{equation}
P_{1,2}^{(l)}=\prod_{j=0,j\neq l}^{2s}\frac{(\vec S_1+\vec
S_2)^2-j(j+1)}{l(l+1)-j(j+1)}.
\end{equation}
The $R_{1,2}^{(s,s)}(u)$ acting on the
$(2s+1)\times(2s+1)$-dimensional tensor space $V_1\otimes V_2$
satisfies the properties:
\begin{eqnarray}
\mbox{Initial condition: } R_{1,2}^{(s,s)}(0)=(2s)!\eta^{2s}{\bf P_{1,2}},\label{R-(s,s)-InitialCondition}\\[2pt]
\mbox{Antisymmetry: }
R_{1,2}^{(s,s)}(-\eta)=(-1)^{2s}(2s+1)!\eta^{2s}P_{1,2}^{(0)}\label{R-(s,s)-Antisymmety},
\end{eqnarray}
where ${\bf P_{1,2}}$ is the permutation operator in the tensor space of two spin-$s$ spaces.

The $R$-matrix (\ref{r111}) of the spin-$s$ Heisenberg spin chain
can be constructed by the fusion procedure \cite{Kulish1,Kulish2,Kulish3,Kirillov1,Kirillov2}. The starting
point is the fundamental spin-$\frac{1}{2}$ $R$-matrix
\begin{eqnarray}
R_{1,2}^{(\frac12,\frac12)}(u)=u+\eta P_{1,2},
\end{eqnarray}
where $P_{1,2}=\frac12 (1+\vec\sigma_1\cdot\vec\sigma_2)$ is the
permutation operator defined in the tensor space of spin-$\frac{1}{2}$ spaces and $\vec\sigma$ is the Pauli matrix. By
taking the fusion in the quantum space, we obtain the
spin-$(\frac{1}{2},s)$ $R$-matrix $R_{1,2}^{(\frac{1}{2},s)}(u)$
defined in the spin-$\frac{1}{2}$ auxiliary space (two-dimensional)
and the spin-$s$ quantum space ($2s+1$-dimensional) as
\begin{eqnarray}\label{R-matrix-(1/2,s)}
R_{1,2}^{(\frac{1}{2},s)}(u)&=&u+\frac{\eta}{2}+\eta \vec{\sigma}_1\cdot \vec S_2\nonumber\\
&=&\left(
\begin{array}{cccc}
u+\frac{\eta}{2}+\eta S_2^z & \eta S_2^-\\
\eta S_2^+ & u+\frac{\eta}{2}-\eta S_2^z
\end{array}
\right),
\end{eqnarray}
where $\vec S$ is the spin-$s$ operator and $S^\pm=S^x\pm iS^y$. The
$R$-matrix (\ref{R-matrix-(1/2,s)}) can also be expressed as
\begin{eqnarray}\label{R-matrix-fusion}
R_{a,\{1,\cdots,2s\}}^{(\frac12,s)}(u)&=&\frac{1}{\prod_{k=1}^{2s-1}(u+(\frac12-s+k)\eta)}\nonumber\\
&\times& P_{\{1,\cdots,2s\}}^{(+)}\prod_{k=1}^{2s}\bigg \{R_{a,k}^{(\frac12,\frac12)}(u+(k-\frac12-s)\eta)\bigg\}P_{\{1,\cdots,2s\}}^{(+)},
\end{eqnarray}
with the product in the order of increasing $k$ from the left to the
right, where $P_{\{1,\cdots,2s\}}^{(+)}$ is the symmetric projector
given by
\begin{equation}
P_{\{1,\cdots,2s\}}^{(+)}=\frac{1}{(2s)!}\prod_{k=1}^{2s}\bigg(\sum_{l=1}^k P_{l,k}\bigg).
\end{equation}
Further more, taking the fusion in the auxiliary space,
the spin-$(j,s)$ $R$-matrix can be given by
\begin{eqnarray}
\fl\quad\quad\quad
R_{\{1,\cdots,2j\},\{1,\cdots,2s\}}^{(j,s)}(u)&=&P_{\{1,\cdots,2s\}}^{(+)}\prod_{k=1}^{2j}
\bigg\{R_{k,\{1,\cdots,2s\}}^{(\frac12,s)}(u+(k-j-\frac12)\eta)\bigg\}P_{\{1,\cdots,2s\}}^{(+)},\nonumber\\
&&\quad\quad j,s=\frac12,1,\frac23,\cdots.
\end{eqnarray}
The spin-$(s_i,s_j)$ $R$-matrix $R_{i,j}^{(s_i,s_j)}(u)$ acting on
the $(2s_i+1)\times(2s_j+1)$-dimensional tensor space $V_i\otimes
V_j$ satisfies the Yang-Baxter equation
\begin{equation}\label{QYBE}
\fl \quad\quad
R_{1,2}^{(s_1,s_2)}(u-v)R_{1,3}^{(s_1,s_3)}(u)R_{2,3}^{(s_2,s_3)}(v)=R_{2,3}^{(s_2,s_3)}(v)R_{1,3}^{(s_1,s_3)}(u)R_{1,2}^{(s_1,s_2)}(u-v).
\end{equation}

The reflection matrix $K^{-(s)}$ of spin-$s$ Heisenberg spin chain
can also be obtained by the fusion procedure developed in \cite{Mezincescu,fusion1,fusion2}
\begin{eqnarray}
K_{\{a\}}^{-(s)}(u)&=&P_{\{a\}}^{(+)}\prod_{k=1}^{2s}\bigg\{\bigg[\prod_{l=1}^{k-1}R_{a_l,a_k}^{(\frac12,\frac12)}(2u+(k+l-2s-1)\eta)\bigg]\nonumber\\
&\times&K_{ak}^{-(\frac12)}(u+(k-s-\frac12)\eta)\bigg\}P_{\{a\}}^{(+)},
\end{eqnarray}
which satisfies the reflection equation \cite{sklya}
\begin{eqnarray}
&&R_{\{a\},\{b\}}^{(j,s)}(u-v)K^{-(j)}_{\{a\}}(u)R^{(s,j)}_{\{b\},\{a\}}(u+v)K^{-(s)}_{\{b\}}(v)\nonumber\\[4pt]
&=&K^{-(s)}_{\{b\}}(v)R^{(j,s)}_{\{a\}\{b\}}(u+v)K^{-(j)}_{\{a\}}(u)R^{(s,j)}_{\{b\}\{a\}}(u-v),\label{RE}
\end{eqnarray}
and ${K_0^-}^{(\frac{1}{2})}(u)$ is the fundamental spin-1/2
reflection matrix given by \cite{deva2,Gho}:
\begin{eqnarray}
{K_0^-}^{(\frac{1}{2})}(u)= \left(
\begin{array}{cc}
p+u  & \varsigma u\\
\varsigma u & p-u
\end{array}
\right)\equiv \left(
\begin{array}{cc}
K_{11}^{-}(u)  & K_{12}^{-}(u) \\
K_{21}^{-}(u)  & K_{22}^{-}(u)
\end{array}
\right),\label{K-1/2}
\end{eqnarray}
where $p$ and $\varsigma$ are two generic boundary parameters. The
corresponding dual reflection matrix $K^{+(s)}(u)$ is thus defined as
\begin{equation}
K^{+(s)}_{\{a\}}(u)=\frac{1}{f^{(s)}(u)}K^{-(s)}_{\{a\}}(-u-\eta)\bigg|_{(p,\varsigma)\rightarrow
(q,-\xi)},
\end{equation}
where $q$ and $\xi$ are two generic boundary parameters and the
normalization operator $f^{(s)}(u)$ is
\begin{eqnarray}
&&f^{(s)}(u)=\prod_{l=1}^{2s-1}\prod_{k=1}^l\bigg[-\phi(2u+(l+k+1-2s)\eta)\bigg],\\[2pt]
&&\phi(u)=(u+\eta)(u-\eta).
\end{eqnarray}
The fundamental spin-1/2 dual
reflection matrix reads
\begin{eqnarray}
{K_0^+}^{(\frac{1}{2})}(u)= \left(
\begin{array}{cc}
q-u-\eta  & \xi(u+\eta)\\
\xi(u+\eta) & q+u+\eta
\end{array}
\right)\equiv \left(
\begin{array}{cc}
K_{11}^{+}(u)  & K_{12}^{+}(u) \\
K_{21}^{+}(u)  & K_{22}^{+}(u)
\end{array}
\right).
\end{eqnarray}

The one-row monodromy matrices for spin-$(j,s)$ are given by
\begin{eqnarray}
&&T_{\{a\}}^{(j,s)}(u)=R_{\{a\},\{b^{[N]}\}}^{(j,s)}(u-\theta_N)\cdots R_{\{a\},\{b^{[1]}\}}^{(j,s)}(u-\theta_1),\\[4pt]
&&\hat
T_{\{a\}}^{(j,s)}(u)=R_{\{b^{[1]}\},\{a\}}^{(s,j)}(u+\theta_N)\cdots
R_{\{b^{[N]}\},\{a\}}^{(s,j)}(u+\theta_N),
\end{eqnarray}
which satisfy the  Yang-Baxter relations
\begin{eqnarray}\label{YBERTT2}
R_{0,0'}^{(j,j)}(u-v)T_0^{(j,s)}(u)T_{0'}^{(j,s)}(v)
=T_{0'}^{(j,s)}(v)T_0^{(j,s)}(u) R_{0,0'}^{(j,j)}(u-v), \\[2pt]
R_{0,0'}^{(j,j)}(u-v)\hat T_0^{(j,s)}(u)\hat T_{0'}^{(j,s)}(v) =\hat
T_{0'}^{(j,s)}(v)\hat T_0^{(j,s)}(u) R_{0,0'}^{(j,j)}(u-v),
\end{eqnarray}
where $\{\theta_j|_j=1,\cdots,N\}$ are some generic inhomogeneity parameters
and $N$ is the number of sites.
Accordingly, the double-row monodromy matrix for spin-$(j,s)$ is defined as
\begin{equation}
\mathscr{U}_0^{(j,s)}(u)=T_0^{(j,s)}(u)K_0^{-(j)}(u)\hat
T_0^{(j,s)}(u),
\end{equation}
which satisfies the reflection equation
\begin{eqnarray}\label{YBERURU2}
&&R_{0,0'}^{(j,j)}(u-v)\mathscr{U}_0^{(j,s)}(u)R_{0',0}^{(j,j)}(u+v)\mathscr{U}_{0'}^{(j,s)}(v)\nonumber \\[2pt]
&=&\mathscr{U}_{0'}^{(j,s)}(v)R_{0',0}^{(j,j)}(u+v)\mathscr{U}_0^{(j,s)}(u)R_{0,0'}^{(j,j)}(u-v).
\end{eqnarray}
The spin-$(j,s)$ transfer matrix is thus defined as
\begin{eqnarray}
t^{(j,s)}(u)=tr_{\{a\}}\bigg\{
K_{\{a\}}^{+(j)}(u)\mathscr{U}_{\{a\}}^{(j,s)}(u)\bigg\}.
\end{eqnarray}
The corresponding Hamiltonian in terms of the transfer
matrix $t^{(s,s)}(u)$ is thus given by
\begin{eqnarray}
H=\frac{\partial}{\partial u}\{\ln
[f^{(s)}(u)\,t^{(s,s)}(u)]\}|_{u=0,\{\theta_j=0\}}.
\end{eqnarray}

From the Yang-Baxter equation (\ref{QYBE}), the reflection equation
(\ref{RE}) and its dual version, one can check that the transfer
matrix with different spectral parameters are mutually commutative
for arbitrary $j,j',s\in\{\frac12,1,\frac23,\cdots\}$
\begin{equation}
[t^{(j,s)},t^{(j',s)}]=0 ,
\end{equation}
which implies that they have common eigenstates. In fact, the transfer matrices
$\{t^{(j,s)}(u)\}$ satisfy the fusion hierarchy relation \cite{fusion1,fusion2}
\begin{eqnarray}\label{FusionHierarchyRelation}
\fl \quad\quad t^{(\frac12,s)}(u)t^{(j-\frac12,s)}(u-j\eta)=t^{(j,s)}(u-(j-\frac12)\eta)+\delta^{(s)}(u)t^{(j-1,s)}(u-(j+\frac12)\eta,\nonumber\\
\quad\quad\quad\quad\quad\quad\quad\quad\quad\quad\quad\quad\quad\quad\quad\quad\quad\quad
j=\frac12,1,\frac32,\cdots,
\end{eqnarray}
with $t^{(0,s)}(u)=id$ and
\begin{eqnarray}
\delta^{(s)}(u)
&=&\frac{(2u-2\eta)(2u+2\eta)}{(2u-\eta)(2u+\eta)}((1+\varsigma^2)u^2-p^2)((1+\xi^2)u^2-q^2)\nonumber\\
&&\times \prod_{l=1}^N (u-\theta_l+(\frac12+s)\eta)(u+\theta_l+(\frac12+s)\eta)\nonumber\\
&&\times \prod_{l=1}^N (u-\theta_l-(\frac12+s)\eta)(u+\theta_l-(\frac12+s)\eta).\nonumber
\end{eqnarray}
With the initial condition (\ref{R-(s,s)-InitialCondition}) of
$R_{1,2}^{(s,s)}(u)$, the hierarchy relation
(\ref{FusionHierarchyRelation}) is closed at the inhomogeneity
points \cite{Cao5}
\begin{eqnarray}\label{t(s,s) operator identity}
\fl \quad\quad t^{(s,s)}(\theta_l)t^{(\frac12,s)}(\theta_l-(\frac12+s)\eta)=\delta^{(s)}(\theta_l+(\frac12-s)\eta)t^{(s-\frac12,s)}(\theta_l+(\frac12+s)\eta),
\nonumber\\
\quad\quad\quad\quad\quad\quad\quad\quad\quad\quad\quad\quad\quad l=1,\cdots,N,
\end{eqnarray}
which together with the crossing symmetry
$t^{(\frac12,s)}(-u-\eta)=t^{(\frac12,s)}(u)$ and the asymptotic
behavior
\begin{eqnarray}
t^{(\frac12,s)}(u)|_{u\rightarrow\infty}=2(\xi\varsigma-1)u^{2N+2}\times
id+\cdots,\label{t(s,s) relation2}
\end{eqnarray}
\begin{eqnarray}
t^{(\frac12,s)}(0)=2pq\prod_{l=1}^N(\theta_l+(\frac12+s)\eta)(-\theta_l+(\frac12+s)\eta)\times
id, \label{t(s,s) relation3}
\end{eqnarray}
allows us to express $\Lambda^{(\frac12,s)}(u)$ , the eigenvalues of $t^{(\frac12,s)}(u)$, in
the following inhomogeneous $T-Q$ formalism \cite{Cao5}
\begin{eqnarray}\label{Lambda}
\fl\quad \quad\Lambda^{(\frac12,s)}(u)=a^{(s)}(u)\frac{Q(u-\eta)}{Q(u)}+d^{(s)}(u)\frac{Q(u+\eta)}{Q(u)}
+cu(u+\eta)\frac{F^{(s)}(u)}{Q(u)},
\end{eqnarray}
where the functions $a^{(s)}(u),d^{(s)}(u),F^{(s)}(u)$ and the
constant $c$ are given by
\begin{eqnarray}
&&a^{(s)}(u)=\frac{2u+2\eta}{2u+\eta}(\sqrt{1+\xi^2}u+q)(\sqrt{1+\varsigma^2}u+p)\nonumber\\
&&\quad\quad\quad\times\prod_{l=1}^N(u-\theta_l+(\frac12+s)\eta)(u+\theta_l+(\frac12+s)\eta),\\
&&d^{(s)}(u)=a^{(s)}(-u-\eta)\\
&&\quad\quad\quad=\frac{2u}{2u+\eta}(\sqrt{1+\xi^2}(-u-\eta)+q)(\sqrt{1+\varsigma^2}(-u-\eta)+p)\nonumber\\
&&\quad\quad\quad\times\prod_{l=1}^N(-u-\theta_l+(-\frac12+s)\eta)(-u+\theta_l+(-\frac12+s)\eta),\\
&&F^{(s)}(u)=\prod_{l=1}^N\prod_{k=0}^{2s}(u-\theta_l+(\frac12-s+k)\eta)(u+\theta_l+(\frac12-s+k)\eta),\\[2pt]
&&c=2(\varsigma\xi-1-\sqrt{1+\varsigma^2}\sqrt{1+\xi^2}).
\end{eqnarray}
The $Q$-function is parameterized as
\begin{eqnarray}
Q(u)&&=\prod_{j=1}^{2sN}(u-\lambda_j)(u+\lambda_j+\eta),
\end{eqnarray}
and the $2sN$ Bethe roots $\{\lambda_j|j=1,\cdots,2sN\}$ should
satisfy the Bethe ansatz equations (BAEs)
\begin{eqnarray}
&&a^{(s)}(\lambda_j)Q(\lambda_j-\eta) +d^{(s)}(\lambda_j)Q(\lambda_j+\eta)
+c\,\lambda_j(\lambda_j+\eta)\,F^{(s)}(\lambda_j)=0, \nonumber \\[2pt]
&&\qquad\qquad j=1,\ldots,2sN.\label{BAE-s-3}
\end{eqnarray}

\section{Gauge transformation}
\setcounter{equation}{0}

Without losing generality, we put $\varsigma=0$ in the following text. For convenience, we introduce the notations
\begin{eqnarray}\label{T0}
&&T_0^{(\frac{1}{2},s)}(u)=\left(
\begin{array}{cc}
A(u) & B(u)\\
C(u) & D(u)
\end{array}
\right),\\
&&\hat T_0^{(\frac{1}{2},s)}(u)=(-1)^N\left(
\begin{array}{cc}
D(-u-\eta) & -B(-u-\eta)\\
-C(-u-\eta) & A(-u-\eta)
\end{array}
\right),
\end{eqnarray}

\begin{equation}\label{DoubleRowMonodromyMatrix}
\mathscr{U}_0^{(\frac{1}{2},s)}(u)=T_0^{(\frac{1}{2},s)}(u)K_0^{-(\frac{1}{2})}(u)\hat
T_0^{(\frac{1}{2},s)}(u) =\left(
\begin{array}{cc}
\mathscr{A}(u) & \mathscr{B}(u)\\
\mathscr{C}(u) & \mathscr{D}(u)
\end{array}
\right).
\end{equation}
Let us introduce the gauge
matrix
\begin{eqnarray}\label{TransU}
U_0=\left(
\begin{array}{cc}
\sqrt{1+\xi^2}-1 &\xi\\
-\sqrt{1+\xi^2}-1 &\xi
\end{array}
\right),\label{gauge}
\end{eqnarray}
with which
$K_0^{+(\frac12)}$-matrix can be diagonalized as
\begin{eqnarray}
\fl \quad\quad\tilde K_0^{+(\frac{1}{2})}(u)&=&U_0 K_0^{+(\frac{1}{2})} (u)U_0^{-1}
=\left(
\begin{array}{cc}
q+\sqrt{1+\xi^2}(u+\eta) & 0\\
0 & q-\sqrt{1+\xi^2}(u+\eta)
\end{array}
\right)\nonumber\\
&=&\left(
\begin{array}{cc}
\tilde K_{11}^{+}(u) & 0\\
0 & \tilde K_{22}^{+}(u)
\end{array}
\right),
\end{eqnarray}
and the gauged $K^{-(\frac12)}$-matrix $\tilde K_0^{-(\frac12)}(u)$ becomes
\begin{eqnarray}
\tilde K_0^{-(\frac{1}{2})}(u)&=&U_0 K_0^{+(\frac{1}{2})}(u) U_0^{-1}
=\left(
\begin{array}{cc}
p-\frac{u}{\sqrt{1+\xi^2}} & -\frac{\sqrt{1+\xi^2}-1}{\sqrt{1+\xi^2}}u\\
-\frac{\sqrt{1+\xi^2}+1}{\sqrt{1+\xi^2}}u & p+\frac{u}{\sqrt{1+\xi^2}}
\end{array}
\right)\nonumber\\
&=&\left(
\begin{array}{cc}
\tilde K_{11}^{-}(u) & \tilde K_{12}^{-}(u)\\
\tilde K_{21}^{-}(u) & \tilde K_{22}^{-}(u)
\end{array}
\right).
\end{eqnarray}

Accordingly, the one-row monodromy matrices under the above gauge transformation read
\begin{eqnarray}
\hspace{-0.5cm}\tilde T_0^{(\frac{1}{2},s)}(u)&=&U_0 T_0^{(\frac{1}{2},s)}(u) U_0^{-1}
=\left(
\begin{array}{cc}
\tilde A(u) & \tilde B(u)\\
\tilde C(u) & \tilde D(u)
\end{array}
\right),\label{GaugedT}\nonumber\\
\hspace{-0.5cm}\tilde{\hat T}_0^{(\frac{1}{2},s)}(u)&=&U_0 \hat T_0^{(\frac{1}{2},s)}(u) U_0^{-1}
=(-1)^N\left(
\begin{array}{cc}
\tilde D(-u-\eta) & -\tilde B(-u-\eta)\\
-\tilde C(-u-\eta) &\tilde A(-u-\eta)
\end{array}
\right).\label{GaugedHatT}
\end{eqnarray}
The double-row monodromy matrix $\tilde{\mathscr{U}}_0^{(\frac{1}{2},s)}(u)$ is gauged to
\begin{eqnarray}\label{TransDoubleRowMonodromyMatrix}
\tilde{\mathscr{U}}_0^{(\frac{1}{2},s)}(u)&=&U_0T_0^{(\frac{1}{2},s)}(u)K_0^{-(\frac{1}{2})}(u)\hat T_0^{(\frac{1}{2},s)}(u)U_0^{-1}\nonumber\\
&=&\tilde T_0^{(\frac{1}{2},s)}(u)\tilde K_0^{-(\frac{1}{2})}(u)\tilde{\hat T}_0^{(\frac{1}{2},s)}(u)
=\left(
\begin{array}{cc}
\tilde{\mathscr{A}}(u) & \tilde{\mathscr{B}}(u)\\
\tilde{\mathscr{C}}(u) & \tilde{\mathscr{D}}(u)
\end{array}
\right),
\end{eqnarray}
which gives the following relations
\begin{eqnarray}
\tilde{\mathscr{A}}(u)&=&(-1)^N\{\tilde K_{11}^{-}(u)\tilde
A(u)\tilde D(-u-\eta)+\tilde K_{21}^{-}(u)\tilde B(u)\tilde
D(-u-\eta)\nonumber\\[4pt]
&&-\tilde K_{12}^{-}(u)\tilde A(u)\tilde C(-u-\eta)-\tilde
K_{22}^{-}(u)\tilde B(u)\tilde C(-u-\eta)\},\label{tildeA}\\[4pt]
\tilde{\mathscr{B}}(u)&=&(-1)^N\{-\tilde K_{11}^{-}(u)\tilde
A(u)\tilde B(-u-\eta)-\tilde K_{21}^{-}(u)\tilde B(u)\tilde
B(-u-\eta)\nonumber\\[4pt]
&&+\tilde K_{12}^{-}(u)\tilde A(u)\tilde A(-u-\eta)+\tilde
K_{22}^{-}(u)\tilde B(u)\tilde A(-u-\eta)\},\label{tildeB}\\[4pt]
\tilde{\mathscr{C}}(u)&=&(-1)^N\{\tilde K_{11}^{-}(u)\tilde
C(u)\tilde D(-u-\eta)+\tilde K_{21}^{-}(u)\tilde D(u)\tilde
D(-u-\eta)\nonumber\\ [4pt]
&&-\tilde K_{12}^{-}(u)\tilde
C(u)\tilde C(-u-\eta)-\tilde K_{22}^{-}(u)\tilde D(u)\tilde
C(-u-\eta)\}, \label{tildeC}\\[4pt]
\tilde{\mathscr{D}}(u)&=&(-1)^N\{-\tilde K_{11}^{-}(u)\tilde
C(u)\tilde B(-u-\eta)-\tilde K_{21}^{-}(u)\tilde D(u)\tilde
B(-u-\eta)\nonumber\\[4pt]
&&+\tilde K_{12}^{-}(u)\tilde C(u)\tilde A(-u-\eta)+\tilde
K_{22}^{-}(u)\tilde D(u)\tilde A(-u-\eta)\}.\label{tildeD}
\end{eqnarray}
The transfer matrix $t^{(\frac12,s)}(u)$ can be expressed as
\begin{eqnarray}
\fl \quad\quad t^{(\frac12,s)}(u) =tr_0(\tilde
K_0^{+(\frac{1}{2})}(u)\tilde{\mathscr{U}}_0^{(\frac{1}{2},s)}(u))=\tilde
K_{11}^{+(\frac{1}{2})}(u)\tilde{\mathscr{A}}(u)+\tilde
K_{22}^{+(\frac{1}{2})}(u) \tilde{\mathscr{D}}(u).
\end{eqnarray}

Thanks to the $SU(2)$-invariance of the $R$-matrix,  the gauged one-row monodromy matrix
also satisfies the relation
\begin{eqnarray}
R_{0,0'}^{(\frac12,\frac12)}(u-v)\tilde T_0^{(\frac{1}{2},s)}(u)\tilde T_{0'}^{(\frac{1}{2},s)}(v)
=\tilde T_{0'}^{(\frac{1}{2},s)}(v)\tilde T_0^{(\frac{1}{2},s)}(u) R_{0,0'}^{(\frac12,\frac12)}(u-v), \nonumber
\end{eqnarray}
which gives rise to the following commutation relations
\begin{eqnarray}
&&\tilde A(u)\tilde B(v)=\frac{u-v-\eta}{u-v}\tilde B(v)\tilde A(u)+\frac{\eta}{u-v}\tilde B(u)\tilde A(v),\\[4pt]
&&\tilde D(u)\tilde B(v)=\frac{u-v+\eta}{u-v}\tilde B(v)\tilde D(u)-\frac{\eta}{u-v}\tilde B(u)\tilde D(v),\\[4pt]
&&\tilde B(u)\tilde D(v)=\frac{u-v+\eta}{u-v}\tilde D(v)\tilde B(u)-\frac{\eta}{u-v}\tilde D(u)\tilde B(v),\\[4pt]
&&\tilde C(u)\tilde A(v)=\frac{u-v+\eta}{u-v}\tilde A(v)\tilde C(u)-\frac{\eta}{u-v}\tilde A(u)\tilde C(v),\\[4pt]
&&\tilde C(u)\tilde D(v)=\frac{u-v-\eta}{u-v}\tilde D(v)\tilde C(u)+\frac{\eta}{u-v}\tilde D(u)\tilde C(v),\\[4pt]
&&[\tilde C(u),\tilde B(v)]=\frac{\eta}{u-v}[\tilde D(u)\tilde
A(v)-\tilde D(v)\tilde A(u)].
\end{eqnarray}
Similarly, the gauged double-row monodromy matrix satisfies
\begin{eqnarray}\label{UYBERURU1}
&&R_{0,0'}^{(\frac{1}{2},\frac12)}
(u-v)\tilde{\mathscr{U}}_0^{(\frac{1}{2},s)}(u)R_{0',0}^{(\frac{1}{2},\frac12)}(u+v)\tilde{\mathscr{U}}_{0'}^{(\frac{1}{2},s)}(v)\nonumber\\[4pt]
&=&\tilde{\mathscr{U}}_{0'}^{(\frac{1}{2},s)}
(v)R_{0',0}^{(\frac{1}{2},\frac12)}(u+v)\tilde{\mathscr{U}}_0^{(\frac{1}{2},s)}(u)R_{0,0'}^{(\frac{1}{2},\frac12)}(u-v),
\end{eqnarray}
which leads to the following commutation relations
\begin{eqnarray}
&&\tilde{\mathscr{C}}(u)\tilde{\mathscr{A}}(v)
=\frac{(u+v)(u-v+\eta)}{(u-v)(u+v+\eta)}\tilde{\mathscr{A}}(v)\tilde{\mathscr{C}}(u)
-\frac{\eta}{u+v+\eta}\tilde{\mathscr{D}}(u)\tilde{\mathscr{C}}(v)\nonumber\\[2pt]
&&\quad\quad\quad\quad\quad
-\frac{(u+v)\eta}{(u-v)(u+v+\eta)}\tilde{\mathscr{A}}(u)\tilde{\mathscr{C}}(v),\label{DoubleRowMonoCommutationCA}\\[2pt]
&&\tilde{\mathscr{D}}(v)\tilde{\mathscr{C}}(u)
=\frac{(u+v)(u-v+\eta)}{(u-v)(u+v+\eta)}\tilde{\mathscr{C}}(u)\tilde{\mathscr{D}}(v)
-\frac{\eta}{u+v+\eta}\tilde{\mathscr{C}}(v)\tilde{\mathscr{A}}(u)\nonumber\\[2pt]
&&\quad\quad\quad\quad\quad
-\frac{(u+v)\eta}{(u-v)(u+v+\eta)}\tilde{\mathscr{C}}(v)\tilde{\mathscr{D}}(u),\label{DoubleRowMonoCommutationDC}\\[2pt]
&&\tilde{\mathscr{A}}(u)\tilde{\mathscr{A}}(v)
=\tilde{\mathscr{A}}(v)\tilde{\mathscr{A}}(u)
+\frac{\eta}{u+v+\eta}\tilde{\mathscr{B}}(v)\tilde{\mathscr{C}}(u)\nonumber\\[2pt]
&&\quad\quad\quad\quad\quad
-\frac{\eta}{u+v+\eta}\tilde{\mathscr{B}}(u)\tilde{\mathscr{C}}(v),\label{DoubleRowMonoCommutationAA}\\[2pt]
&&\tilde{\mathscr{D}}(u)\tilde{\mathscr{D}}(v)
=\tilde{\mathscr{D}}(v)\tilde{\mathscr{D}}(u)
+\frac{\eta}{u+v+\eta}\tilde{\mathscr{C}}(v)\tilde{\mathscr{B}}(u)\nonumber\\[2pt]
&&\quad\quad\quad\quad\quad
-\frac{\eta}{u+v+\eta}\tilde{\mathscr{C}}(u)\tilde{\mathscr{B}}(v),\label{DoubleRowMonoCommutationDD}\\[2pt]
&&\tilde{\mathscr{D}}(u)\tilde{\mathscr{A}}(v)
=\tilde{\mathscr{A}}(v)\tilde{\mathscr{D}}(u)
-\frac{\eta(u+v+2\eta)}{(u-v)(u+v+\eta)}\tilde{\mathscr{B}}(u)\tilde{\mathscr{C}}(v)\nonumber\\[2pt]
&&\quad\quad\quad\quad\quad
+\frac{\eta(u+v+2\eta)}{(u-\nu)(u+v+\eta)}\tilde{\mathscr{B}}(v)\tilde{\mathscr{C}}(u),\label{DoubleRowMonoCommutationDA}\\[4pt]
&&[\tilde{\mathscr{C}}(u),\tilde{\mathscr{C}}(v)]=[\tilde{\mathscr{B}}(u),\tilde{\mathscr{B}}(v)]=0.\label{DoubleRowMonoCommutationCCBB}
\end{eqnarray}

\section{Orthogonal Basis}
\setcounter{equation}{0}

In order to obtain the orthogonal basis of the Hilbert space, we
first introduce the reference state. For general spin-$s$ cases, the
gauged $\tilde R_{0,n}^{(\frac12,s)}$ is
\begin{eqnarray}
\tilde R_{0,n}^{(\frac12,s)}(u)&=&U_0 R_{0,n}^{(\frac12,s)}(u)
U_0^{-1}\equiv \left(
\begin{array}{cc}
\tilde r_{11}(u) & \tilde r_{12}(u)\\
\tilde r_{21}(u) & \tilde r_{22}(u)
\end{array}
\right),
\end{eqnarray}
where \begin{eqnarray}
\fl \quad\quad&&\tilde r_{21}(u)=-\frac{1}{2\xi\sqrt{1+\xi^2}}[2\xi(\sqrt{1+\xi^2}+1)\eta S_n^z+(\sqrt{1+\xi^2}+1)^2\eta S_n^--\xi^2\eta S_n^+],\\[4pt]
\fl \quad\quad&&\tilde
r_{12}(u)=-\frac{1}{2\xi\sqrt{1+\xi^2}}[2\xi(\sqrt{1+\xi^2}-1)\eta
S_n^z-(\sqrt{1+\xi^2}-1)^2\eta S_n^-+\xi^2\eta S_n^+].
\end{eqnarray}
We introduce a set of local states $\{|\tilde s_a\rangle_n
=\sum_kc_k^{(a)}|k\rangle_n, a=1,\cdots,2s+1, k=-s,\cdots, s, n=1,
\cdots, N\}$, where $\{|k\rangle_n, k=-s,\cdots, s\}$  form the eigenstates of $S_n^z$, i.e., $S_n^z|k\rangle_n=k|k\rangle_n$.
The coefficients $\{c_k^{(1)}\}$ are determined by the constraint
\begin{eqnarray}
\tilde r_{21}|\tilde s_1\rangle_n=0,\nonumber
\end{eqnarray}
which gives the coefficients of $|\tilde s_1\rangle_n$ as
\begin{eqnarray}
&&c_{-s+j}^{(1)}=\frac{\sqrt{2s(2s-1)\cdots(2s-j+1)}}{\sqrt{j!}(\sqrt{1+\xi^2}+1)^{j-2}}\xi^j,\quad j=0,\cdots,2s,
\end{eqnarray}
The coefficients $\{c_k^{(a)},a=2,\cdots,2s+1\}$ are determined by the
condition
\begin{equation}
|\tilde s_a\rangle_n=f{(\eta)}\tilde r_{12}|\tilde s_{a-1}\rangle_n,
\end{equation}
which gives rise to the values of $\{c_k^{(2s+1)}\}$ as ($f(\eta)$ is an irrelevant normalization factor)
\begin{eqnarray}
&&  c_{-s+j}^{(2s+1)}=(-1)^j\frac{\sqrt{2s(2s-1)\cdots(2s-j+1)}}{\sqrt{j!}(\sqrt{1+\xi^2}-1)^{j-2}}\xi^j,\quad j=0,\cdots,2s,
\end{eqnarray}
The reference states $\{|\tilde s_a\rangle_n, n=1, \cdots, N \}$
satisfy the following orthogonal relations
\begin{equation}
_j\langle \tilde s_a|\tilde
s_b\rangle_j=\delta_{a,b},\quad a,b=1,2,\cdots,2s+1, \quad
j=1,\dots,N.
\end{equation}

We introduce the product state $|\Omega\rangle=\bigotimes_{n=1}^N
|\tilde s_1\rangle_n$ and $\langle \bar
\Omega|=\bigotimes_{n=1}^N{_n}\langle\tilde s_{2s+1}|$, which are the
eigenstates of the operators $\tilde A(u)$ and $\tilde D(u)$
\begin{eqnarray}
&&\tilde A(u)|\Omega\rangle=a(u)|\Omega\rangle,\quad\tilde D(u)|\Omega\rangle=d(u)|\Omega\rangle,\quad \tilde C(u)|\Omega\rangle=0, \label{rad}\\[2pt]
&&\langle\bar\Omega|\tilde A(u)=d(u)\langle\bar\Omega|,\quad
\langle\bar\Omega|\tilde D(u)=a(u)\langle\bar\Omega|,\quad
\langle\bar\Omega|\tilde C(u)=0,\label{lad}
\end{eqnarray}
with the corresponding eigenvalues
\begin{equation}\label{audu}
a(u)=\prod_{l=1}^N (u-\theta_l+(\frac12+s)\eta),\quad\quad d(u)=\prod_{l=1}^N (u-\theta_l+(\frac12-s)\eta).
\end{equation}
Denoting $\beta'_l\equiv \theta_l-(\frac12+s)\eta$ and $\beta_l\equiv
\theta_l-(\frac12-s)\eta$, we have $a(\beta'_l)=0$ and
$d(\beta_l)=0$. From the equation (\ref{tildeC}), we find that the product
state $|\Omega\rangle$ and $\langle \bar \Omega|$ are also the
eigenstates of the operator $\tilde{\mathscr{C}}(u)$
\begin{eqnarray}
\tilde{\mathscr{C}}(u)|\Omega\rangle=(-1)^N {\tilde K}^-_{21}(u)d(u)d(-u-\eta)|\Omega\rangle,\\[2pt]
\langle \bar \Omega|\tilde{\mathscr{C}}(u)=(-1)^N {\tilde
K}^-_{21}(u)a(u)a(-u-\eta)\langle\bar\Omega|.
\end{eqnarray}

Noting the fact that
$[\tilde{\mathscr{C}}(u),\tilde{\mathscr{C}}(v)]=0$, the eigenstates of $\tilde{\mathscr{C}}(u)$ can form a basis of the
Hilbert space in the sense of Sklyanin's separation of variables \cite{skl1,skl2,skl3}. Let us introduce the following states
\begin{eqnarray}
\fl\quad\quad|\beta^{(\alpha_{1})}_{1},\cdots,\beta^{(\alpha_{N})}_{N}\rangle=\prod_{j=1}^N\prod_{k_{j}
=0}^{\alpha_{j}-1}\tilde{\mathscr{A}}(\beta_{j}-k_{j}\eta)|\Omega\rangle,\quad\quad
\alpha_{j}=0,1,\cdots,2s, \label{RightBasis} \\
\fl\quad\quad\langle\beta'^{(\alpha_{1})}_{1},\cdots,\beta'^{(\alpha_{N})}_{N}|=\langle\bar\Omega|\prod_{j=1}^N\prod_{k_{j}
=0}^{\alpha_{j}-1}\tilde{\mathscr{D}}(-\beta'_{j}-(k_{j}+1)\eta),\quad\quad
\alpha_{j}=0,1,\cdots,2s.\label{LeftBasis}
\end{eqnarray}
It should be noted that the products of
$\tilde{\mathscr{A}}(\beta_{j}-k_{j}\eta)$ in Eq.(\ref{RightBasis})
are ordered by decreasing $k_j$ while
$\tilde{\mathscr{D}}(-\beta'_{j}-(k_{j}+1)\eta)$ in
(\ref{LeftBasis}) are ordered by increasing $k_j$ from left to
right. Using the commutation relations
(\ref{DoubleRowMonoCommutationCA})-(\ref{DoubleRowMonoCommutationDA}),
we conclude that Eq.(\ref{RightBasis}) and Eq.(\ref{LeftBasis}) are
eigenstates of $\tilde{\mathscr{C}}(u)$
\begin{eqnarray}
\hspace{-0.5cm}&&\tilde{\mathscr{C}}(u)|\beta^{(\alpha_{1})}_{1},\cdots,\beta^{(\alpha_{N})}_{N}\rangle=
h(u,\{\beta^{(\alpha_{1})}_{1},\cdots,\beta^{(\alpha_{N})}_{N}\})
|\beta^{(\alpha_{1})}_{1},\cdots,\beta^{(\alpha_{N})}_{N}\rangle,\label{CAeigenstates}
\\[4pt]
\hspace{-0.5cm}&&\langle
\beta'^{(\alpha_{1})}_{1},\cdots,\beta'^{(\alpha_{N})}_{N}|\tilde{\mathscr{C}}(u)=
\bar
h(u,\{\beta'^{(\alpha_{1})}_{1},\cdots,\beta'^{(\alpha_{N})}_{N}\})
\langle\beta'^{(\alpha_{1})}_{1},\cdots,\beta'^{(\alpha_{N})}_{N}|,\label{DCeigenstates}
\end{eqnarray}
with the eigenvalues
\begin{eqnarray}
\hspace{-0.5cm}h(u,\{\beta^{(\alpha_{1})}_{1},\cdots,\beta^{(\alpha_{N})}_{N}\})
&=&(-1)^N\tilde K^-_{21}(u)d(-u-\eta)d(u)\nonumber\\[2pt]
&&\times\prod_{j=1}^N
\frac{(u-\beta_{j}+\alpha_{j}\eta)(u+\beta_{j}+\eta-\alpha_{j}\eta)}
{(u-\beta_{j})(u+\beta_{j}+\eta)},
\label{CCAEigenstates} \\[2pt]
\hspace{-0.5cm}\bar
h(u,\{\beta'^{(\alpha_{1})}_{1},\cdots,\beta'^{(\alpha_{N})}_{N}\})
&=&(-1)^N\tilde K^-_{21}(u)a(-u-\eta)a(u)\nonumber\\[2pt]
&&\times\prod_{j=1}^N
\frac{(u-\beta'_{j}-\alpha_j\eta)(u+\beta'_{j}+\eta+\alpha_j\eta)}
{(u-\beta'_{j})(u+\beta'_{j}+\eta)}. \label{CDCEigenstates}
\end{eqnarray}
By using the commutation relations
(\ref{DoubleRowMonoCommutationCA})-(\ref{DoubleRowMonoCommutationDA})
and Eqs.(\ref{CAeigenstates})-(\ref{DCeigenstates}), we can prove
that the order of the product of $\tilde{\mathscr{A}}$
$(\tilde{\mathscr{D}})$ with respect to different $\beta_{j}$ $(\beta'_j)$ in Eq.(\ref{RightBasis})
(Eq.(\ref{LeftBasis})) is changeable, while the order of that with
the same $\beta_{j}$ $(\beta'_j)$ can not be changed. The right states
given by Eq.(\ref{RightBasis}) (the left states given by
Eq.(\ref{LeftBasis})) form a complete and orthogonal basis of the
Hilbert space. Therefore, the eigenstates of the transfer matrices can be
decomposed as a unique linear combination of the basis vectors.

\section{The scalar product}
\setcounter{equation}{0}

For convenience, we introduce
\begin{eqnarray}
\overline{\tilde{\mathscr{D}}}(u)=\tilde{\mathscr{D}}(u)-\frac{\eta}{2u+\eta}\tilde{\mathscr{A}}(u).
\end{eqnarray}
The transfer matrix $t^{(\frac12,s)}(u)$ can be expressed as
\begin{eqnarray}
t^{(\frac12,s)}(u)=\left[\tilde{K}^+_{11}(u)+\frac{\eta}{2u+\eta}\tilde{K}^+_{22}(u)\right]\tilde{\mathscr{A}}(u)
+\tilde{K}^+_{22}(u)\overline{\tilde{\mathscr{D}}}(u).
\end{eqnarray}
Let $\langle \Psi|$ be an eigenstate of the transfer matrix of
$t^{(\frac12,s)}(u)$, namely,
\begin{equation}
\langle \Psi|t^{(\frac12,s)}(u)=\langle \Psi|\Lambda^{(\frac12,s)}
(u),
\end{equation}
where the eigenvalue $\Lambda^{(\frac12,s)} (u)$ is given by the
inhomogeneous $T-Q$ relation (\ref{Lambda}).

Now let us evaluate the scalar product
\begin{equation}
F(\alpha_1,\cdots,\alpha_N)=\langle
\Psi|\beta^{(\alpha_{1})}_{1},\cdots,\beta^{(\alpha_{N})}_{N}\rangle,
\end{equation}
by calculating the quantity $\langle
\Psi|t^{(\frac12,s)}(\beta_{{n}}-m\eta)|\beta^{(\alpha_{1})}_{1},\cdots,\beta^{(\alpha_{n}=m)}_{n},\cdots,\beta^{(\alpha_{N})}_{N}\rangle$.
Acting $t^{(\frac12,s)}(\beta_{{n}}-m\eta)$ to the left and to the right
alternately, we obtain
\begin{eqnarray}\label{Fn11}
\fl\qquad &&\Lambda^{(\frac12,s)}(\beta_{{n}}-m\eta)F(\alpha_1,\cdots,\alpha_{n}=m,\cdots,\alpha_N) \nonumber \\[4pt]
\fl\qquad &=&\left[\tilde K^+_{11}(\beta_{{n}}-m\eta)+\frac{\eta\tilde K^+_{22}(\beta_{{n}}-m\eta)}{2\beta_{{n}}-(2m-1)\eta}\right]F(\alpha_1,\cdots,\alpha_{n}=m+1,\cdots,\alpha_N)\nonumber\\
\fl \qquad&&+\tilde
K^+_{22}(\beta_{{n}}-m\eta)\langle\Psi|\overline{\tilde{\mathscr{D}}}(\beta_{{n}}-m\eta)
|\beta^{(\alpha_{1})}_{1},\cdots,\beta^{(\alpha_n=m)}_{n},\cdots,\beta^{(\alpha_{N})}_{N}\rangle.\label{Frelation}
\end{eqnarray}
From Eqs.(\ref{tildeA}) and (\ref{tildeD}), we have the following
relations
\begin{eqnarray}
\fl \qquad \tilde{\mathscr{A}}(u)|\Omega\rangle=(-1)^N\bigg\{\tilde
K_{11}^-(u)a(u)d(-u-\eta)
|\Omega\rangle+\tilde K_{21}^-(u)d(-u-\eta)\tilde{B}(u)|\Omega\rangle\bigg\},\\[4pt]
\fl \qquad  \overline{\tilde{\mathscr{D}}}(u)|\Omega\rangle=(-1)^N\bigg\{\frac{(2u+\eta)\tilde K_{22}^-(u)-\eta\tilde K_{11}^-(u)}{2u+\eta}d(u)a(-u-\eta)|\Omega\rangle\nonumber\\[2pt]
\quad-\frac{2u+2\eta}{2u+\eta}\tilde
K_{21}^-(u)d(u)\tilde{B}(-u-\eta)|\Omega\rangle\bigg\}.
\end{eqnarray}
It is easy to check
\begin{eqnarray}
\qquad\qquad\overline{\tilde{\mathscr{D}}}(\beta_j)|\Omega\rangle=0,\quad
j=1,\cdots,N,{\label{D0}}
\end{eqnarray}
which allows us to write
$F(\alpha_1,\cdots,\alpha_{n}=1,\cdots,\alpha_N)$ as
\begin{eqnarray}
\fl \qquad &&F(\alpha_1,\cdots,\alpha_{n}=1,\cdots,\alpha_N)\nonumber\\[4pt]
\fl \qquad &=&\frac{(2\beta_n+\eta)\Lambda^{(\frac12,s)}(\beta_n)}
{(2\beta_n+\eta)\tilde{K}^+_{11}(\beta_n)+\eta\tilde{K}^+_{22}(\beta_n)}F(\alpha_1,\cdots,\alpha_{n}=0,\cdots,\alpha_N)\nonumber\\[4pt]
\fl \qquad &=&(-1)^N(p+\beta_{n})a(\beta_{n}) d(-\beta_{n}-\eta)
\frac{Q(\beta_{n}-\eta)}{Q(\beta_{n})}F(\alpha_1,\cdots,\alpha_{n}=0,\cdots,\alpha_N).\label{recursive
relation1}
\end{eqnarray}
Based on the properties of quantum determinant \cite{IK} (for a detailed description, see \cite{Book}),
\begin{eqnarray}
\fl \qquad{\rm{Det}}_q\{\tilde{T}^{(\frac12,s)}(u)\}&=&\tilde{A}(u-\eta)\tilde{D}(u)-\tilde{C}(u-\eta)\tilde{B}(u)\nonumber\\[4pt]
\fl \qquad&=&\tilde{D}(u-\eta)\tilde{A}(u)-\tilde{B}(u-\eta)\tilde{C}(u),\\[2pt]
\fl \qquad{\rm{Det}}_q\{\tilde{\hat{T}}^{(\frac12,s)}(u)\}&=&\tilde{A}(-u)\tilde{D}(-u-\eta)-\tilde{B}(-u)\tilde{C}(-u-\eta)\nonumber\\[4pt]
\fl
\qquad&=&\tilde{D}(-u)\tilde{A}(-u-\eta)-\tilde{C}(-u)\tilde{B}(-u-\eta),
\end{eqnarray}
and the commutation relations
\begin{eqnarray}
\tilde{A}(u)\tilde{B}(u-\eta)&=&\tilde{B}(u)\tilde{A}(u-\eta),\quad \tilde{C}(u)\tilde{D}(u-\eta)=\tilde{D}(u)\tilde{C}(u-\eta),\\[4pt]
\tilde{D}(u-\eta)\tilde{B}(u)&=&\tilde{B}(u-\eta)\tilde{D}(u),\quad
\tilde{A}(u-\eta)\tilde{C}(u)=\tilde{C}(u-\eta)\tilde{A}(u),
\end{eqnarray}
we find that the following relation holds
\begin{eqnarray}
&&\overline{\tilde{\mathscr{D}}}(u-\eta)\tilde{\mathscr{A}}(u)-\frac{2u}{2u-\eta}\tilde{\mathscr{B}}(u-\eta)\tilde{\mathscr{C}}(u)\nonumber\\[4pt]
&=&\frac{1}{2u-\eta}{\rm{Det}}_q\{\tilde{\mathscr{U}}^{(\frac12,s)}(u)\}\nonumber\\[4pt]
&=&\frac{2u-2\eta}{2u-\eta}(p^2-u^2)a(u)d(-u-\eta)a(-u)d(u-\eta).\label{DET}
\end{eqnarray}
According to Eqs.(\ref{CAeigenstates}) and (\ref{CCAEigenstates}),
we know
\begin{eqnarray}
\tilde{\mathscr{C}}(\beta_n-\alpha_n\eta)|\beta^{(\alpha_{1})}_{1},\cdots,\beta^{(\alpha_{n})}_{n},\cdots,\beta^{(\alpha_{N})}_{N}\rangle=0.
\label{ttt}
\end{eqnarray}
Using the relations (\ref{DET}) and (\ref{ttt}), we obtain
\begin{eqnarray}
\fl \quad&&\overline{\tilde{\mathscr{D}}}(\beta_{{n}}-m\eta)
|\beta^{(\alpha_{1})}_{1},\cdots,\beta^{(\alpha_n=m)}_{n},\cdots,\beta^{(\alpha_{N})}_{N}\rangle\nonumber\\[4pt]
\fl \quad&=&\frac{2\beta_n-2m\eta}{2\beta_n-(2m-1)\eta}\left\{p^2-[\beta_n-(m-1)\eta]^2\right\}a(\beta_n-(m-1)\eta)d(-\beta_n+(m-2)\eta)\nonumber\\[4pt]
\fl  \quad&&\times a(-\beta_n+(m-1)\eta)d(\beta_n-m\eta)|\beta^{(\alpha_{1})}_{1},\cdots,\beta^{(\alpha_n=m-1)}_{n},\cdots,\beta^{(\alpha_{N})}_{N}\rangle,\nonumber\\[4pt]
\fl &&\quad m=1,\cdots,2s.\label{D eigenstate}
\end{eqnarray}
Substituting Eq.(\ref{D eigenstate}) into (\ref{Frelation}), we
obtain the recursive relations about $F(\alpha_1,\cdots,\alpha_N)$
\begin{eqnarray}
\fl\quad &&\Lambda^{(\frac12,s)}(\beta_{{n}}-m\eta)F(\alpha_1,\cdots,\alpha_{n}=m,\cdots,\alpha_N)\nonumber\\[4pt]
\fl\quad &=&\left[\tilde K^+_{11}(\beta_{{n}}-m\eta)+\frac{\eta\tilde K^+_{22}(\beta_{{n}}-m\eta)}{2\beta_{{n}}-2m\eta+\eta}\right]F(\alpha_1,\cdots,\alpha_{n}=m+1,\cdots,\alpha_N)\nonumber\\
\fl \quad&&+ \frac{2\beta_n-2m\eta}{2\beta_n-(2m-1)\eta}{\tilde K}^+_{22}(\beta_{{n}}-m\eta)\left\{p^2-[\beta_n-(m-1)\eta]^2\right\}a(\beta_n-(m-1)\eta)\nonumber\\[4pt]
\fl  \quad&&\times d(-\beta_n+(m-2)\eta)a(-\beta_n+(m-1)\eta)d(\beta_n-m\eta)\nonumber\\[4pt]
\fl \quad &&\times
F(\alpha_1,\cdots,\alpha_{n}=m-1,\cdots,\alpha_N),\quad
m=1,\cdots,2s-1.\label{recursive relation2}
\end{eqnarray}
The initial condition (\ref{recursive relation1}) and the recursive
relations (\ref{recursive relation2}) give rise to
\begin{eqnarray}
F(\alpha_1,\cdots,\alpha_N)
=\prod_{j=1}^N\prod_{k_j=0}^{\alpha_j-1}(-1)^N(p+\beta_{j}-k_j\eta)\nonumber\\
\qquad\times a(\beta_{j}-k_j\eta) d(-\beta_{j}+(k_j-1)\eta)
\frac{Q(\beta_{j}-(k_j+1)\eta)}{Q(\beta_{j}-k_j\eta)}F_0,\label{UniRecursiveRelation}
\end{eqnarray}
where $F_0=\langle \Psi|\Omega\rangle$ is an overall scalar factor.

\section{The inner product $\langle 0|\beta^{(\alpha_{1})}_{1},\cdots,\beta^{(\alpha_{N})}_{N}\rangle$}
\setcounter{equation}{0}

The definition of the one-row monodonomy matrix $T_0(u)$ implies
\begin{equation}
\langle0|A(u)=a(u)\langle0|,\quad \langle0|D(u)=d(u)\langle0|,\quad
\langle0|B(u)=0,
\end{equation}
where the functions $a(u)$ and $d(u)$ are given by Eq.(\ref{audu}),
$\langle0|={_1}\langle s|\otimes\cdots\otimes _N\langle s|$. The double-row monodromy matrix
(\ref{DoubleRowMonodromyMatrix}) acting on the state $\langle0|$
gives
\begin{eqnarray}
\langle 0|\mathscr{A}(u)&=&(-1)^N K^-_{11}(u)a(u)d(-u-\eta)\langle 0|,\\[4pt]
\langle 0|\mathscr{D}(u)&=&(-1)^N \frac{\eta}{2u+\eta}K^-_{11}(u)a(u)d(-u-\eta)\langle 0| \nonumber\\[4pt]
&&+(-1)^N \frac{(2u+\eta)K^-_{22}(u)-\eta K^-_{11}(u)}{2u+\eta}a(-u-\eta)d(u)\langle 0|,\\[4pt]
\langle 0|\mathscr{B}(u)&=&0,\\[4pt]
\langle 0|\mathscr{C}(u)&=&(-1)^N \frac{2u}{2u+\eta}K^-_{11}(u)d(-u-\eta)\langle 0|C(u)\nonumber\\[4pt]
&&+(-1)^N \frac{-(2u+\eta)K^-_{22}(u)+\eta
K^-_{11}(u)}{2u+\eta}d(u)\langle 0|C(-u-\eta).
\end{eqnarray}

Notice that the following relations hold
\begin{eqnarray}
\tilde{\mathscr{A}}(u)&=&\frac{1}{2\xi\sqrt{1+\xi^2}}\bigg\{\xi(\sqrt{1+\xi^2}-1)\mathscr{A}(u)
+\xi^2\mathscr{C}(u)\nonumber\\[2pt]
&&+\xi^2\mathscr{B}(u)+\xi(1+\sqrt{1+\xi^2})\mathscr{D}(u)\bigg\},\\[2pt]
\tilde{\mathscr{C}}(u)&=&\frac{1}{2\xi\sqrt{1+\xi^2}}\bigg\{-\xi(1+\sqrt{1+\xi^2})\mathscr{A}(u)
-(1+\sqrt{1+\xi^2})^2\mathscr{B}(u)\nonumber\\[2pt]
&&+\xi^2\mathscr{C}(u)+\xi(1+\sqrt{1+\xi^2})\mathscr{D}(u)\bigg\},\\[2pt]
\tilde{\mathscr{D}}(u)&=&\frac{1}{2\xi\sqrt{1+\xi^2}}\bigg\{\xi(1+\sqrt{1+\xi^2})\mathscr{A}(u)
-\xi^2\mathscr{C}(u)\nonumber\\[2pt]
&&-\xi^2\mathscr{B}(u)+\xi(-1+\sqrt{1+\xi^2})\mathscr{D}(u)\bigg\}.
\end{eqnarray}
The relation $\langle 0|\tilde {\mathscr{C}}(\beta_{{n}}-(m-1)\eta)
|\beta^{(\alpha_{1})}_{1},\cdots,\beta^{(\alpha_{n}=m-1)}_{n},\cdots,\beta^{(\alpha_{N})}_{N}\rangle=0$
gives rise to
\begin{eqnarray}\label{relation1}
&&\quad\langle 0| {\mathscr{C}}(\beta_{{n}}-(m-1)\eta)
|\beta^{(\alpha_{1})}_{1},\cdots,\beta^{(\alpha_{n}=m-1)}_{n},\cdots,\beta^{(\alpha_{N})}_{N}\rangle
\nonumber\\[4pt]
&&=[1+\sqrt{1+\xi^2}]\xi^{-1}\langle 0| \{{\mathscr{A}}(\beta_{{n}}-(m-1)\eta)-{\mathscr{D}}(\beta_{{n}}-(m-1)\eta)\}\nonumber\\[4pt]
&&\quad\times|\beta^{(\alpha_{1})}_{1},\cdots,\beta^{(\alpha_{n}=m-1)}_{n},\cdots,\theta^{(\alpha_{N})}_{N}\rangle.
\end{eqnarray}
With the help of Eq.(\ref{relation1}), we have
\begin{eqnarray}\label{VacBasisN2}
&&\quad\langle 0|\beta^{(\alpha_{1})}_{1},\cdots,\beta^{(\alpha_{n}=m)}_{n},\cdots,\beta^{(\alpha_{N})}_{N}\rangle
\nonumber\\[4pt]
&&=\langle 0|\tilde{\mathscr{A}}(\beta_{{n}}-(m-1)\eta)
|\beta^{(\alpha_{1})}_{1},\cdots,\beta^{(\alpha_{n}=m-1)}_{n},\cdots,\beta^{(\alpha_{N})}_{N}\rangle\nonumber\\[4pt]
&&=(-1)^N K^-_{11}(\beta_{{n}}-(m-1)\eta)a(\beta_{n}-(m-1)\eta)d(-\beta_{n}+(m-2)\eta)\nonumber\\[4pt]
&&\quad\times\langle
0|\beta^{(\alpha_{1})}_{1},\cdots,\beta^{(\alpha_{n}=m-1)}_{n},\cdots,\beta^{(\alpha_{N})}_{N}\rangle,\nonumber
\end{eqnarray}
which induces the solution
\begin{eqnarray}
\langle
0|\beta^{(\alpha_{1})}_{1},\cdots,\beta^{(\alpha_{N})}_{N}\rangle
&=&\prod_{j=1}^N\prod_{k_j=0}^{\alpha_j-1}(-1)^N K^-_{11}(\beta_{{j}}-k_j\eta)\nonumber\\[2pt]
&&\times a(\beta_{j}-k_j\eta)d(-\beta_{j}+(k_j-1)\eta)\langle
0|\Omega\rangle.
\end{eqnarray}

\section{Bethe States}
\setcounter{equation}{0}

We introduce the following left Bethe states
\begin{equation}\label{LeftBetheStateN}
\langle \lambda_1,\cdots,\lambda_{2sN}|=\langle 0|\bigg\{\prod_{j=1}^{2sN}\frac{\tilde {\mathscr{C}}(\lambda_j)}
{(-1)^N\tilde K^-_{21}(\lambda_j)d(\lambda_j)d(-\lambda_j-\eta)}\bigg\}.\label{Bethe-state-1}
\end{equation}
%
The relations (\ref{CAeigenstates}) and (\ref{VacBasisN2}) imply
that
\begin{eqnarray}
&&\langle\lambda_1,\cdots,\lambda_{2sN}|\beta^{(\alpha_{1})}_{1},\cdots,\beta^{(\alpha_{N})}_{N}\rangle\nonumber\\[2pt]
&=&\prod_{j=1}^N\prod_{k_j=0}^{\alpha_j-1}(-1)^N(p+\beta_{j}-k_j\eta)a(\beta_{j}-k_j\eta)\nonumber\\
&&\times
 d(-\beta_{j}+(k_j-1)\eta)
\frac{Q(\beta_{j}-(k_j+1)\eta)}{Q(\beta_{j}-k_j\eta)}\langle0|\Omega\rangle,\nonumber
\end{eqnarray}
which is consistent with Eq.(\ref{UniRecursiveRelation}). Therefore,
we conclude that the Bethe states given by
Eq.(\ref{LeftBetheStateN}) are the eigenstates of the transfer
matrix $t^{(\frac12,s)}(u)$, provided that the Bethe roots $\{\lambda_j|j=1,\cdots,2sN\}$ satisfy the BAEs (\ref{BAE-s-3}). With a similar procedure, we can
construct the right Bethe states of the transfer matrices as
\begin{equation}\label{RightBetheStateN}
| \lambda_1,\cdots,\lambda_{2sN}\rangle=\bigg\{\prod_{j=1}^{2sN}\frac{\tilde {\mathscr{B}}(\lambda_j)}
{(-1)^N\tilde K^-_{12}(\lambda_j)a(\lambda_j)a(-\lambda_j-\eta)}\bigg\}|0\rangle, \label{Bethe-state-2}
\end{equation}
with $|0\rangle=|s\rangle_1\otimes\cdots\otimes |s\rangle_N$.

From the definitions (\ref{gauge}) of the gauge matrix, it is clear that both the reference state $|0\rangle$ (or $\langle 0|$) and the generator
$\tilde {\mathscr{B}}(u)$ (or $\tilde {\mathscr{C}}(u)$) have well-defined homogeneous limits: $\{\theta_j\rightarrow 0\}$.  This implies that
the homogeneous limit of the Bethe state (\ref{Bethe-state-2}) exactly gives rise to the corresponding Bethe state of the homogeneous
spin-$s$ chain with generic open  boundaries, where the associated $T-Q$ relation and BAEs are given by (\ref{Lambda}) and (\ref{BAE-s-3}) with $\{\theta_j=0\}$.

\section{Conclusions}

In conclusion, the Bethe-type eigenstates of the integrable spin-$s$
Heisenberg chain with generic open boundary condition are constructed based on the inhomogeneous $T-Q$ relation. It is shown
that the resulting Bethe states have well-defined homogeneous limits.
The method developed in this paper provides a possible way to construct Bethe-type eigenstates of high-level integrable models with generic boundary conditions. It should be remarked that a generic scalar product $\langle\Psi|\prod_{j=1}^M\tilde {\mathscr{B}}(u_j)|0\rangle$, which is relevant to the form factors, can be expressed easily as a linear combination of $F(\alpha_1,\cdots,\alpha_N)$.

\section*{Acknowledgments}

The financial supports from the National Natural Science Foundation
of China (Grant Nos. 11375141, 11374334, 11434013 and 11425522), the
National Program for Basic Research of MOST (973 project under grant
No. 2011CB921700), BCMIIS and the Strategic Priority Research
Program of the CAS are gratefully acknowledged.

\end{document}